# Decentralized Anti-coordination
# Through Multi-agent Learning

**Ludek Cigler**                                    LUDEK.CIGLER@EPFL.CH
**Boi Faltings**                                    BOI.FALTINGS@EPFL.CH
*Artificial Intelligence Laboratory*
*Ecole Polytechnique Fédérale de Lausanne*
*CH-1015 Lausanne, Switzerland*

## Abstract

To achieve an optimal outcome in many situations, agents need to choose distinct actions from one another. This is the case notably in many resource allocation problems, where a single resource can only be used by one agent at a time. How shall a designer of a multi-agent system program its identical agents to behave each in a different way?

From a game theoretic perspective, such situations lead to undesirable Nash equilibria. For example consider a resource allocation game in that two players compete for an exclusive access to a single resource. It has three Nash equilibria. The two pure-strategy NE are efficient, but not fair. The one mixed-strategy NE is fair, but not efficient. Aumann's notion of correlated equilibrium fixes this problem: It assumes a correlation device that suggests each agent an action to take.

However, such a "smart" coordination device might not be available. We propose using a randomly chosen, "stupid" integer coordination signal. "Smart" agents learn which action they should use for each value of the coordination signal.

We present a multi-agent learning algorithm that converges in polynomial number of steps to a correlated equilibrium of a channel allocation game, a variant of the resource allocation game. We show that the agents learn to play for each coordination signal value a randomly chosen pure-strategy Nash equilibrium of the game. Therefore, the outcome is an efficient correlated equilibrium. This CE becomes more fair as the number of the available coordination signal values increases.

## 1. Introduction

In many situations, agents have to coordinate their actions in order to use some limited resource: In communication networks, a channel might be used by only one agent at a time. When driving a car, an agent prefers to choose a road with less traffic, i.e. the one chosen by a smaller number of other agents. When bidding for one item in several simultaneous auctions, an agent prefers the auction with less participants, because this will usually lead to a lower price. Such situations require agents to take each a different decision. However, all the agents are identical, and the problem each one of them face is the same. How can they learn to behave differently from everyone else?

Second problem arises when agents have common preferences over which action they want to take: In the communication networks problem, every agent prefers to transmit over being quiet. In the traffic situation, agents might all prefer the shorter path. But in order to achieve an efficient allocation, it is necessary precisely for some agents to stay quiet, to take the longer path. How can we achieve that those agents are not exploited? How can the agents learn to alternate, taking the longer road on one day, while taking the shorter road the next day?

A central coordinator who possesses complete information about agents' preferences and about the available resources can easily recommend each agent which action to take. However, such an omniscient central coordinator is not always available. Therefore, we would like to be able to use a distributed scheme. We consider scenarios where the same agents try to use the same set of resources repeatedly, and can use the history of the past interactions to learn to coordinate the access to the resources in the future.

In particular, consider the problem of radio channel access. In this problem, $N$ agents try to transmit over $C$ non-overlapping channels. Fully decentralized schemes, such as ALOHA (Abramson, 1970), can only achieve a throughput of $\frac{1}{e} \approx 37\%$, i.e. any transmission only succeeds with probability $\frac{1}{e} \approx 0.37$. More complex schemes, such as those based on distributed constraint optimization (Cheng, Raja, Xie, & Howitt, 2009), can reach a throughput close to 100%. By throughput, we mean the probability of successful transmission on a given channel. However, the messages necessary to implement these schemes create an overhead that eliminates part of their benefits.

In this paper we propose instead to use a simple signal that all agents can observe and that ergodically fluctuates. This signal could be a common clock, radio broadcasting on a specified frequency, the decimal part of a price of a certain stock at a given time, etc. Depending on this "stupid" signal, "smart" agents can then learn to take a different action for each of its value. We say that the signal is stupid, because it doesn't have anything to do with the game – it is the agents who give it a meaning themselves by acting differently for each value of the coordination signal.

Let's look at the simplest example of a channel access problem: one where 2 agents try to transmit over one shared channel. We can model this situation as a game in normal form. Agents can choose between two actions: to stay *quiet* ($Q$) or to *transmit* ($T$). Only one agent may transmit successfully at a time. If an agent transmits alone, she receives a positive payoff. If an agent does not transmit on the channel, her payoff is 0. If *both* agents try to transmit on the channel at the same time, their transmissions fail and they incur a cost $c$.

The payoff matrix of the game looks as follows:

|   | $Q$ | $T$ |
|---|---|---|
| $Q$ | 0, 0 | 0, 1 |
| $T$ | 1, 0 | $-c, -c$ |

Such a game has two pure-strategy Nash equilibria (NE), where one player stays quiet and the other one transmits. It has also one mixed-strategy NE, where each player stays quiet with probability $\frac{1}{c+1}$. The two pure-strategy NE are efficient, in that they maximize the social welfare, but they are not fair: Only one player gets the full payoff, even though the game is symmetric. The mixed-strategy NE is fair, but not efficient: The expected payoff of both players is 0.

As such, the Nash equilibria of the game are rather undesirable: they are either efficient or fair, but not both at the same time. In his seminal paper, Aumann (1974) proposed the notion of *correlated equilibrium* that fixes this problem. A correlated equilibrium (CE) is a probability distribution over the joint strategy profiles in the game. A correlation device samples this distribution and recommends an action for each agent to play. The probability distribution is a CE if agents do not have an incentive to deviate from the recommended action.

In the simple game described above, there exists a CE that is both fair and socially efficient: the correlation device samples from each of the two pure-strategy NE with probability $\frac{1}{2}$ and then recommends the players which NE they should play. This corresponds to an authority that tells each player whether to stay quiet or transmit.

Correlated equilibria have several nice properties: They are easier to find – for a succinct representation of a game, in polynomial time, see (Papadimitriou & Roughgarden, 2008). Also, every Nash equilibrium is a correlated equilibrium. Also, any convex combination of two correlated equilibria is a correlated equilibrium. However, a "smart" correlation device that randomizes over joint strategy profiles might not always be available.

It is possible to achieve a correlated equilibrium without the actual correlation device. Assume that the game is played repeatedly, and that agents can observe the history of actions taken by their opponents. They can learn to predict the future action (or a distribution of future actions) of the opponents. These predictions need to be *calibrated*, that is, the predicted probability that an agent $i$ will play a certain action $a_j$ should converge to the actual frequency with which agent $i$ plays action $a_j$. Agents always play an action that is the best response to their predictions of opponents' actions. Foster and Vohra (1997) showed that in such a case, the play converges to a set of correlated equilibria.

However, in their paper, Foster and Vohra did not provide a specific learning rule to achieve a certain CE. Furthermore, their approach requires that every agent were able to observe actions of every other opponent. If this requirement is not met, convergence to a correlated equilibrium is not guaranteed anymore.

In this paper, we focus on a generalization of the simple channel allocation problem described above. There are $N$ agents who always have some data to transmit, and there are $C$ channels over which they can transmit. We assume that $N \geq C$. Access to a channel is slotted, that is, all agents are synchronized so that they start transmissions at the same time. Also, all transmissions must have the same length. If more than one agent attempts to transmit over a single channel, a collision occurs and none of the transmissions are successful. An unsuccessful transmission has a cost for the agent, since it has to consume some of its (possibly constrained) power for no benefit. Not transmitting does not cost anything.

We assume that agents only receive binary feedback. If they transmitted some data, they find out whether their transmission was successful. If they did not transmit, they can choose some channel to observe. They receive information whether the observed channel was free or not.

When described as a normal-form game, this problem has several efficient (but unfair) pure-strategy Nash equilibria, where a group of $C$ agents gets assigned all the channels. The remaining $N - C$ agents get stranded. It has also a fair but inefficient mixed-strategy NE, where agents choose the transmission channels at random. As in the example of a resource allocation game above, there exists a correlated equilibrium that is efficient and fair.

The "stupid" coordination signal introduced in this paper helps the agents to learn to play a (potentially) different efficient outcome for each of its value. This way, they can reach an efficient allocation while still preserving some level of fairness.

The main contributions of this work are the following:

- We propose a learning strategy for agents in the channel allocation game that, using minimal information, converges in polynomial time to a randomly chosen efficient pure-strategy Nash equilibrium of the game.

- We show that when the agents observe a common discrete correlation signal, they learn to play such an efficient pure-strategy NE for each signal value. The result is a correlated equilibrium that is increasingly fair as the number of available signals $K$ increases.

- We experimentally evaluate how sensitive the algorithm is to a player population that is dynamic, i.e. when players leave and enter the system. We also evaluate the algorithm's resistance to noise, be it in the feedback players receive or in the coordination signal they observe.

The channel allocation algorithm proposed in this paper has been implemented by Wang, Wu, Hamdi, and Ni (2011) in a real-world wireless network setting. They showed how the wireless devices can use all use the actual data which are being transmitted as a coordination signal. That way, they were able to achieve 2-3× throughput gain compared to random access protocols such as ALOHA.

It is worth noting that in this work, our focus is on reaching a correlated and fair outcome, provided that the agents are willing to cooperate. In situations where using resources costs nothing, a self-interested agent could stubbornly keep using it. Everyone else will then be better off not trying to access the resource. This is sometimes called the *"watch out I am crazy"* or *"bully"* strategy (Littman & Stone, 2002).

In order to prevent this kind of behavior, we would need to make sure that in order to use a resource, an agent has to pay some cost. Such a cost may already be implicit in the problem, such as the fact that wireless transmission costs energy, or it may be imposed by external payments. In our recent work (Cigler & Faltings, 2012), we show how this leads to equilibria where rational agents are indifferent between accessing the resource and yielding, and how these equilibria implement our allocation policy for rational agents. We consider this issue to be beyond the scope of this paper and refer the reader to our other work for a deeper analysis.

The rest of the paper is organized as follows: In Section 2, we give some basic definitions from game theory and the theory of Markov chains which we will use throughout the paper. In Section 3, we present the algorithm agents use to learn an action for each possible correlation signal value. In Section 4 we prove that such an algorithm converges to an efficient correlated equilibrium in polynomial time in the number of agents and channels. We show that the fairness of the resulting equilibria increases as the number of signals $K$ increases in Section 5. Section 6 highlights experiments that show the actual convergence rate and fairness. We also show how the algorithm performs in case the population is changing dynamically. In Section 7 we present some related work from game theory and cognitive radio literature, and Section 8 concludes.

## 2. Preliminaries

In this section, we will introduce some basic concepts of game theory and of the theory of Markov chains that we are going to use throughout the paper.

### 2.1 Game Theory

Game theory is the study of interactions among independent, self-interested agents. An agent who participates in a game is called a *player*. Each player has a utility function associated with each state of the world. Self-interested players take actions so as to achieve a state of the world that maximizes their utility. Game theory studies and attempts to predict the behaviour, as well as the final outcome of such interactions. Leyton-Brown and Shoham (2008) give a more complete introduction to game theory.

The basic way to represent a strategic interaction (*game*) is using the so-called *normal form*.

**Definition 1.** A finite, $N$-person *normal-form game* is a tuple $(\mathbf{N}, \mathcal{A}, u)$, where

- $\mathbf{N}$ is a set of $N$ players;

- $\mathcal{A} = \mathcal{A}_1 \times \mathcal{A}_2 \times \ldots \times \mathcal{A}_N$, where $\mathcal{A}_i$ is a set of actions available to player $i$. Each vector $\mathbf{a} = (a_1, a_2, \ldots, a_N) \in \mathcal{A}$ is called an *action profile*;

- $u = (u_1, u_2, \ldots, u_N)$, where $u_i : \mathcal{A} \to \mathbb{R}$ is a *utility function* for player $i$ that assigns each action vector a certain utility (payoff).

When playing a game, players have to select their *strategy*. A *pure* strategy $\sigma_i$ for player $i$ selects only one action $a_i \in \mathcal{A}_i$. A vector of pure strategies for each player $\sigma = (\sigma_1, \sigma_2, \ldots, \sigma_N)$ is called a *pure strategy profile*. A *mixed* strategy selects a probability distribution over the entire action space, i.e. $\sigma_i \in \Delta(\mathcal{A}_i)$. A *mixed strategy profile* is a vector of mixed strategies for each player.

**Definition 2.** We say that a mixed strategy $\sigma_i^*$ of player $i$ is a *best response* to the strategy profile of the opponents $\sigma_{-i}$ if for any strategy $\sigma_i'$,

$$u_i(\sigma_i^*, \sigma_{-i}) \geq u_i(\sigma_i', \sigma_{-i})$$

One of the basic goals of game theory is to predict an outcome of a strategic interaction. Such outcome should be stable – therefore, it is usually called an *equilibrium*. One requirement for an outcome to be an equilibrium is that none of the players has an incentive to change their strategy, i.e. all players play their best-response to the strategies of the others. This defines perhaps the most important equilibrium concept, the Nash equilibrium:

**Definition 3.** A strategy profile $\sigma = (\sigma_1, \sigma_2, \ldots, \sigma_N)$ is a *Nash equilibrium* (NE) if for every player $i$, her strategy $\sigma_{-i}$ is a best response to the strategies of the others $\sigma_{-i}$.

As essential as Nash equilibria are, they have several disadvantages. First, they may be hard to find: Chen and Deng (2006) show that finding NE is *PPAD-complete*. Second, there might be multiple Nash equilibria, as shown in the example in Section 1. Third, the most efficient NE may not be the most fair one, even in a symmetric game. We give the formal definition of the *correlated equilibrium* which fixes some of these issues:

**Definition 4.** Given an $N$-player game $(\mathbf{N}, \mathcal{A}, u)$, a *correlated equilibrium* is a tuple $(v, \pi, \mu)$, where $v$ is a tuple of random variables $v = (v_1, v_2, \ldots, v_N)$ whose domains are $D = (D_1, D_2, \ldots, D_N)$, $\pi$ is a joint probability distribution over $v$, $\mu = (\mu_1, \mu_2, \ldots, \mu_N)$ is a vector of mappings $\mu_i : D_i \mapsto \mathcal{A}_i$, and for each player $i$ and every mapping $\mu_i' : D_i \mapsto \mathcal{A}_i$ it is the case that

$$\sum_{d \in D} \pi(d) u_i \left( \mu_1(d_1), \mu_2(d_2), \ldots, \mu_N(d_N) \right) \geq \sum_{d \in D} \pi(d) u_i \left( \mu_1'(d_1), \mu_2'(d_2), \ldots, \mu_N'(d_N) \right).$$

## 2.2 Markov Chains

The learning algorithm we propose and analyze in this paper can be described as a randomized algorithm. In a randomized algorithm, some of its steps depend on the value of a random variable. One useful technique to analyze randomized algorithms is to describe its execution as a *Markov chain*.

A Markov chain is a random process with the *Markov property*. A random process is a collection of random variables; usually it describes the evolution of some random value over time. A process has a Markov property if its state (or value) in the next time step depends exclusively on its value in the previous step, and not on the values further in the past. We can say that the process is *memoryless*. If we imagine the execution of a randomized algorithm as a finite-state automaton with non-deterministic steps, it is easy to see how its execution maps to a Markov chain.

The formal definition of a Markov chain is as follows:

**Definition 5.** (Norris, 1998) Let $I$ be a countable set. Each $i \in I$ is called a *state* and $I$ is called the *state space*. We say that $\lambda = (\lambda_i : i \in I)$ is a *measure* on $I$ if $0 \leq \lambda_i < \infty$ for all $i \in I$. If in addition the *total mass* $\sum_{i \in I} \lambda_i$ equals 1, then we call $\lambda$ a *distribution*. We work throughout with a probability space $(\Omega, \mathcal{F}, \mathbb{P})$. Recall that a *random variable* $X$ with values in $I$ is a function $X : \Omega \to I$. Suppose we set

$$\lambda_i = \Pr(X = i) = \Pr \left( \{ \omega \in \Omega : X(\omega) = i \} \right).$$

Then $\lambda$ defines a distribution, the distribution of $X$. We think of $X$ as modelling a random state that takes value $i$ with probability $\lambda_i$.

We say that a matrix $P = (p_{ij} : i, j \in I)$ is *stochastic* if every row $(p_{ij} : j \in I)$ is a distribution.

We say that $(X_t)_{t \geq 0}$ is a *Markov chain* with *initial distribution* $\lambda$ and a *transition matrix* $P$ if

1. $X_0$ has distribution $\lambda$;

2. for $t \geq 0$, conditional on $X_t = i$, $X_{t+1}$ has distribution $(p_{ij} : j \in I)$ and is independent of $X_0, X_1, \ldots, X_{t-1}$.

More explicitly, the conditions state that, for $t \geq 0$ and $i_0, \ldots, i_{t+1} \in I$,

1. $\Pr(X_0 = i_0) = \lambda_{i_0}$;

2. $\Pr(X_{t+1} = i_{t+1} | X_0 = i_0, \ldots, X_t = i_t) = p_{i_t i_{t+1}}$.

**Theorem 1.** *Let $A$ be a set of states. The vector of hitting probabilities $h^A = (h_i^A : i \in \{0, 1, \ldots, N\})$ is the minimal non-negative solution to the system of linear equations*

$$h_i^A = \begin{cases} 1 & \text{for } i \in A \\ \sum_{j \in \{0,1,\ldots,N\}} p_{ij} h_j^A & \text{for } i \notin A \end{cases}$$

Intuitively, the hitting probability $h_i^A$ is the probability that when the Markov chain starts in state $i$, it will ever reach some of the states in $A$.

One property of randomized algorithms that we are particularly interested in is its convergence. If we have a set of states $A$ where the algorithm has converged, we can define the time it takes to reach any state in the set $A$ from any other state of the corresponding Markov chain as the *hitting time*:

**Definition 6.** (Norris, 1998) Let $(X_t)_{t \geq 0}$ be a Markov chain with state space $I$. The *hitting time* of a subset $A \subset I$ is a random variable $H^A : \Omega \to \{0, 1, \ldots\} \cup \{\infty\}$ given by

$$H^A(\omega) = \inf\{t \geq 0 : X_t(\omega) \in A\}$$

Specifically, we are interested in the *expected* hitting time of a set of states $A$, given that the Markov chain starts in an initial state $X_0 = i$. We will denote this quantity

$$k_i^A = \mathbb{E}_i(H^A).$$

In general, the expected hitting time of a set of states $A$ can be found by solving a system of linear equations.

**Theorem 2.** *The vector of expected hitting times $k^A = E(H^A) = (k_i^A : i \in I)$ is the minimal non-negative solution to the system of linear equations*

$$\begin{cases} k_i^A = 0 & \text{for } i \in A \\ k_i^A = 1 + \sum_{j \notin A} p_{ij} k_j^A & \text{for } i \notin A \end{cases} \tag{1}$$

Convergence to an absorbing state may not be guaranteed for a general Markov chain. To calculate the probability of reaching an absorbing state, we can use the following theorem (Norris, 1998):

**Theorem 3.** *Let $A$ be a set of states. The vector of hitting probabilities $h^A = (h_i^A : i \in \{0, 1, \ldots, N\})$ is the minimal non-negative solution to the system of linear equations*

$$h_i^A = \begin{cases} 1 & \text{for } i \in A \\ \sum_{j \in \{0,1,\ldots,N\}} p_{ij} h_j^A & \text{for } i \notin A \end{cases}$$

Solving the systems of linear equations in Theorems 2 and 3 analytically might be difficult for many Markov chains though. Fortunately, when the Markov chain has only one absorbing state $i = 0$, and it can only move from state $i$ to $j$ if $i \geq j$, we can use the following theorem to derive an upper bound on the expected hitting time, proved by Rego (1992):

**Theorem 4.** *Let $A = \{0\}$. If*

$$\forall i \geq 1 : E(X_{t+1} | X_t = i) < \frac{i}{\beta}$$

*for some $\beta > 1$, then*

$$k_i^A < \lceil \log_\beta i \rceil + \frac{\beta}{\beta - 1}$$

## 3. Learning Algorithm

In this section, we describe the algorithm that the agents use to learn a correlated equilibrium of the channel allocation game.

Let us denote the space of available correlation signals $\mathcal{K} := \{0, 1, \ldots, K-1\}$, and the space of available channels $\mathcal{C} := \{1, 2, \ldots, C\}$. Assume that $C \leq N$, that is there are more agents than channels (the opposite case is easier). An agent $i$ has a strategy $f_i : \mathcal{K} \to \{0\} \cup \mathcal{C}$ that she uses to decide which channel she will access at time $t$ when she receives a correlation signal $k_t$. When $f_i(k_t) = 0$, the agent does not transmit at all for signal $k_t$. The agent stores its strategy simply as a table.

Each agent adapts her strategy as follows:

1. In the beginning, for each $k_0 \in \mathcal{K}$, $f_i(k_0)$ is initialized uniformly at random from $\mathcal{C}$. That is, every agent picks a random channel to transmit on, and no agent will monitor other channels.

2. At time $t$:

   - If $f_i(k_t) > 0$, the agent tries to transmit on channel $f_i(k_t)$.
   - If otherwise $f_i(k_t) = 0$, the agent chooses a random channel $m_i(t) \in \mathcal{C}$ that she will monitor for activity.

3. Subsequently, the agent observes the outcome of its choice: if the agent transmitted on some channel, she observes whether the transmission was successful. If it was, the agent will keep her strategy unchanged. If a collision occurred, the agent sets $f_i(k_t) := 0$ with probability $p$. With probability $1 - p$, the strategy remains the same.

4. If the agent did not transmit, she observes whether the channel $m_i(t)$ she monitored was free. If that channel was free, the agent sets $f_i(k_t) := m_i(t)$ with probability 1. If the channel was not free, the strategy $f_i$ remains the same.

## 4. Convergence

An important property of the learning algorithm is if, and how fast it can converge to a pure-strategy Nash equilibrium of the channel allocation game for every signal value. The algorithm is randomized. Therefore, instead of analyzing its worst-case behavior (that may be arbitrarily bad), we will analyze its expected number of steps before convergence.

### 4.1 Convergence for $C = 1, K = 1$

For single channel and single coordination signal, we prove the following theorem:

**Theorem 5.** *For $N$ agents and $C = 1, K = 1, 0 < p < 1$, the expected number of steps before the allocation algorithm converges to a pure-strategy Nash equilibrium of the channel allocation game is $O\left(\frac{1}{p(1-p)} \log N\right)$.*

To prove the convergence of the algorithm, it is useful to describe its execution as a Markov chain.

When $N$ agents compete for a single signal value, a state of the Markov chain is a vector from $\{0, 1\}^N$ that denotes which agents are attempting to transmit. For the purpose of the convergence proof, it is only important how *many* agents are trying to transmit, not which agents. This is because the probability with which the agents back-off is the same for everyone. Therefore, we can describe the algorithm execution using the following chain:

**Definition 7.** A Markov chain describing the execution of the allocation algorithm for $C = 1, K = 1, 0 < p < 1$ is a chain whose state at time $t$ is $X_t \in \{0, 1, \ldots, N\}$, where $X_t = j$ means that $j$ agents are trying to transmit at time $t$.

The transition probabilities of this chain look as follows:

$$\Pr\left(X_{t+1} = N | X_t = 0\right) = 1 \qquad \text{(restart)}$$
$$\Pr\left(X_{t+1} = 1 | X_t = 1\right) = 1 \qquad \text{(absorbing)}$$
$$\Pr\left(X_{t+1} = j | X_t = i\right) = \binom{i}{j} p^{i-j} (1-p)^j \quad i > 1, j \leq i$$

All the other transition probabilities are 0. This is because when there are some agents transmitting on some channel, no other agent will attempt to access it.

The probability $\Pr\left(X_{t+1} = N | X_t = 0\right)$ is equal to 1 because once the channel becomes free ($X_t = 0$), agents will spot this and time $t + 1$, everyone will transmit (therefore the chain will be in state $X_{t+1} = N$). The probability $\Pr\left(X_{t+1} = 1 | X_t = 1\right)$ is equal to one because once a single agent successfully transmits on the channel, she will keep transmitting forever after, and no other agent will attempt to transmit there. Finally, the probability $\Pr\left(X_{t+1} = j | X_t = i\right)$ expresses the fact that when $X_t = i$ ($i$ agents transmit at time $t$), the probability that an agent who transmitted at time $t$ will keep transmitting at time $t + 1$ with probability $1 - p$.

We are interested in the number of steps it will take this Markov chain to first arrive at state $X_t = 1$ given that it started in state $X_0 = N$. This would mean that the agents converged to a setting where only one of them is transmitting, and the others are not. Definition 6 defined the *hitting time* which describes this quantity.

We will show the expected value $E[h_1]$ of the hitting time of state $X_t = 1$ (and by corollary, prove Theorem 5) in the following steps:

1. We show the expected hitting time for a set of states $A = \{0, 1\}$ (Lemma 6)

2. We then show probability that the Markov chain enters state 1 before entering state 0, when it starts from state $i > 1$ (Lemma 7)

3. Finally, using the *law of iterated expectations*, we combine the two lemmas to show the expected hitting time of state 1.

**Lemma 6.** *Let $A = \{0, 1\}$. The expected hitting time of the set of states $A$ in the Markov chain described in Definition 7 is $O\left(\frac{1}{p} \log N\right)$.*

*Proof.* We will first prove that the expected hitting time of a set $A' = \{0\}$ in a slightly modified Markov chain is $O\left(\frac{1}{p} \log N\right)$.

Let us define a new Markov chain $(Y_t)_{t \geq 0}$ with the following transition probabilities:

$$\Pr(Y_{t+1} = 0 | Y_t = 0) = 1 \qquad \text{(absorbing)}$$

$$\Pr(Y_{t+1} = j | Y_t = i) = \binom{i}{j} p^{i-j}(1-p)^j \quad j \geq 0, i \geq 1$$

Note that the transition probabilities are the same as in the chain $(X_t)_{t \geq 0}$, except for states 0 and 1. From state 1 there is a positive probability of going into state 0, and state 0 is now absorbing. Clearly, the expected hitting time of the set $A' = \{0\}$ in the new chain is an upper bound on the expected hitting time of set $A = \{0, 1\}$ in the old chain. This is because any path that leads into state 0 in the new chain either does not go through state 1 (so it happened with the same probability in the old chain), or goes through state 1, so in the old chain it would stop in state 1 (but it would be one step shorter).

If the chain is in state $Y_t = i$, the next state $Y_{t+1}$ is drawn from a binomial distribution with parameters $(i, 1-p)$. The expected next state is therefore

$$E(Y_{t+1} | Y_t = i) = i(1-p)$$

We can therefore use the Theorem 4 with $\beta := \frac{1}{1-p}$ to derive that for $A' = \{0\}$, the hitting time is:

$$k_i^{A'} < \left\lceil \log_{\frac{1}{1-p}} i \right\rceil + \frac{1}{p} \approx O\left(\frac{1}{p} \log i\right)$$

that is also an upper bound on $k_i^A$ for $A = \{0, 1\}$ in the old chain. $\qquad \square$

**Lemma 7.** *The probability $h_i$ that the Markov chain defined in Definition 7 enters state 1 before entering state 0, when started in any state $i > 1$, is greater than $1 - p$.*

*Proof.* Calculating the probability that the chain $X$ enters state 1 before state 0 is equal to calculating the *hitting probability*, i.e. the probability that the chain ever enters a given state, for a modified Markov chain where the probability of staying in state 0 is $\Pr(X_{t+1} = 0 | X_t = 0) = 1$. For a set of states $A$, let us denote $h_i^A$ the probability that the Markov chain starting in state $i$ ever enters some state in $A$. To calculate this probability, we can use Theorem 3. For the modified Markov chain that cannot leave neither state 0 nor state 1, computing $h_i^A$ for $A = 1$ is easy, since the matrix of the system of linear equations is lower triangular.

We'll show that $h_i \geq q = 1 - p$ for $i > 1$ using induction. The first step is calculating $h_i$ for $i \in \{0, 1, 2\}$.

$$h_0 = 0$$
$$h_1 = 1$$
$$h_2 = (1-p)^2 h_2 + 2p(1-p)h_1 + p^2 h_0$$
$$= \frac{2p(1-p)}{1-(1-p)^2} = \frac{2(1-p)}{2-p} \geq 1-p.$$

Now, in the induction step, derive a bound on $h_i$ by assuming $h_j \geq q = 1 - p$ for all $j < i, j \geq 2$.

$$h_i = \sum_{j=0}^{i} \binom{i}{j} p^{i-j}(1-p)^j h_j$$

$$\geq \sum_{j=0}^{i} \binom{i}{j} p^{i-j}(1-p)^j q - ip^{i-1}(1-p)(q-h_1) - p^i h_0$$

$$= q - ip^{i-1}(1-p)(q-1) \geq q = 1-p.$$

This means that no matter which state $i \geq 2$ the Markov chain starts in, it will enter into state 1 earlier than into state 0 with probability at least $1-p$. $\qquad \square$

We can now finish the proof of the bound on the expected hitting time of state 1. We will use the law of iterated expectations:

**Theorem 8.** *(Billingsley, 2012) Let $X$ be a random variable satisfying $E(|X|) < \infty$ and $Y$ another random variable on the same probability space. Then*

$$E[X] = E\left[E[X|Y]\right],$$

*i.e., the expected value of $X$ is equal to the conditional expected value of $X$ given $Y$.*

Let $h_1$ be the random variable corresponding to the hitting time of state 1. Define a random variable $Z$ denoting the number of passes through state 0 our Markov chain makes before it reaches state 1. From Theorem 8, we get:

$$E[h_1] = E[E[h_1|Z]].$$

In Lemma 6, we have shown the expected number of steps $h_A$ before the Markov chain reaches the set of states $A = \{0, 1\}$. We can write

$$E[E[h_1|Z]] = E\left[\sum_{i=1}^{Z} h_A\right] = E[Z \cdot h_A] = h_A \cdot E[Z].$$

From Lemma 7 we know that the probability that the chain passes through state 1 before passing through 0 is greater than $1-p$. Therefore, we can say that $E[Z] \leq E[Z']$ where $Z'$ is a random variable distributed according to a geometric distribution with success probability $1-p$. Then

$$E[h_1] = h_A \cdot E[Z] \leq h_A \cdot E[Z'] = \frac{h_A}{1-p} = O\left(\frac{1}{p(1-p)} \log N\right).$$

This concludes the proof of Theorem 5.

We have just shown that the expected time for convergence of our algorithm is finite, and polynomial. What is the probability that the algorithm converge *in a finite number of steps* to an absorbing state? The following theorem shows that since the expected hitting time of the absorbing state is finite, this probability is 1.

**Theorem 9.** *Let $h_1$ be the hitting time of the state $X_t = 1$ in the Markov chain from Definition 7. Then*

$$\Pr(h_1 \text{ is finite}) = 1.$$

*Proof.* From the Markov inequality, we know that since $h_1 \geq 0$,

$$\Pr(h_1 \geq \alpha) \leq \frac{E[h_1]}{\alpha}.$$

Therefore,

$$\Pr(h_1 \text{ is finite}) = 1 - \lim_{\alpha \to \infty} \Pr(h_1 \geq \alpha) \geq 1 - \lim_{\alpha \to \infty} \frac{E[h_1]}{\alpha} = 1.$$

□

This means that our algorithm converges *almost surely* to a Nash equilibrium of the channel allocation game.

## 4.2 Convergence for $C \geq 1, K = 1$

**Theorem 10.** *For $N$ agents and $C \geq 1, K = 1$, the expected number of steps before the learning algorithm converges to a pure-strategy Nash equilibrium of the channel allocation game is $O\left(C\frac{1}{1-p}\left[\frac{1}{p}\log N + C\right]\right)$.*

*Proof.* In the beginning, in at least one channel, there can be at most $N$ agents who want to transmit. It will take on average $O\left(\frac{1}{p}\log N\right)$ steps to get to a state when either 1 or 0 agents transmit (Lemma 6). We will call this period a *round*.

If all the agents backed off, it will take them on average at most $C$ steps before some of them find an empty channel. We call this period a *break*.

The channels might oscillate between the "round" and "break" periods in parallel, but in the worst case, the whole system will oscillate between these two periods.

For a single channel, it takes on average $O\left(\frac{1}{1-p}\right)$ oscillations between these two periods before there is only one agent who transmits in that channel. For $C \geq 1$, it takes on average $O\left(C\frac{1}{1-p}\right)$ steps between "round" and "break" before all channels have only one agent transmitting. Therefore, it will take on average $O\left(C\frac{1}{1-p}\left[\frac{1}{p}\log N + C\right]\right)$ steps before the system converges. □

## 4.3 Convergence for $C \geq 1, K \geq 1$

To show what is the convergence time when $K > 1$, we will use a more general problem. Imagine that there are $K$ identical instances of the same Markov chain. We know that the original Markov chain converges from any initial state to an absorbing state in expected time $T$. Now imagine a more complex Markov chain: In every step, it selects uniformly at random one of the $K$ instances of the original Markov chain, and executes one step of that instance. What is the time $T_{all}$ before all $K$ instances converge to their absorbing states?

This is an extension of the well-known *Coupon collector's problem* (Feller, 1968). The following theorem (Gast, 2011, Thm. 4) shows an upper bound on the expected number of steps after which all the $K$ instances of the original Markov chain converge:

**Theorem 11.** *(Gast, 2011) Let there be $K$ instances of the same Markov chain that is known to converge to an absorbing state in expectation in $T$ steps. If we select randomly one Markov chain instance at a time and allow it to perform one step of the chain, it will take on average $E[T_{all}] \leq TK \log K + 2TK + 1$ steps before all $K$ instances converge to their absorbing states.*

For arbitrary $C \geq 1, K \geq 1$, the following theorem follows from Theorems 10 and 11:

**Theorem 12.** *For $N$ agents and $C \geq 1, K \geq 1, 0 < p < 1, 0 < q < 1$, the expected number of steps before the learning algorithm converges to a pure-strategy Nash equilibrium of the channel allocation game for every $k \in \mathcal{K}$ is*

$$O\left( (K \log K + 2K) C \frac{1}{1-p} \left[ C + \frac{1}{p} \log N \right] + 1 \right).$$

Aumann (1974) shows that any Nash equilibrium is a correlated equilibrium, and any convex combination of correlated equilibria is a correlated equilibrium. We also know that all the pure-strategy Nash equilibria that the algorithm converges to are efficient: there are no collisions, and in every channel for every signal value, some agent transmits. Therefore, we conclude the following:

**Theorem 13.** *The learning algorithm defined in Section 3 converges in expected polynomial time (with respect to $K$, $C$, $\frac{1}{p}$, $\frac{1}{1-p}$ and $\log N$) to an efficient correlated equilibrium of the channel allocation game.*

## 5. Fairness

Agents decide their strategy independently for each value of the coordination signal. Therefore, every agent has an equal chance that the game converges to an equilibrium that is favorable to her. If the agent can transmit in the resulting equilibrium for a given signal value, we say that the agent *wins* the slot. For $C$ available channels and $N$ agents, an agent wins a given slot with probability $\frac{C}{N}$ (since no agent can transmit in two channels at the same time).

We analyse the fairness of our algorithm *after* it has converged to a correlated equilibrium. While algorithms which do not converge to an absorbing state (such as ALOHA), need to be analysed for all the intermediate states of their execution, we believe that our approach is justified by the fact that our algorithm converges relatively quickly, in polynomial time.

We can describe the number of signals for which an agent $i$ wins some channel as a random variable $X_i$. This variable is distributed according to a binomial distribution with parameters $\left(K, \frac{C}{N}\right)$.

As a measure of fairness, we use the *Jain index* (Jain, Chiu, & Hawe, 1984). The advantage of Jain index is that it is continuous, so that a resource allocation that is strictly more fair has higher Jain index (unlike measures which only assign binary values, such as whether at least half of the agents access some resource). Also, Jain index is independent of the population size, unlike measures such as the standard deviation of the agent allocation.

For a random variable $X$, the Jain index is the following:

$$J(X) = \frac{(E[X])^2}{E[X^2]}$$

When $X$ is distributed according to a binomial distribution with parameters $(K, \frac{C}{N})$, its first and second moments are

$$E[X] = K \cdot \frac{C}{N}$$

$$E[X^2] = \left(K \cdot \frac{C}{N}\right)^2 + K \cdot \frac{C}{N} \cdot \frac{N-C}{N},$$

so the Jain index is

$$J(X) = \frac{C \cdot K}{C \cdot K + (N - C)}. \tag{2}$$

For the Jain index it holds that $0 < J(X) \leq 1$. An allocation is considered fair if $J(X) = 1$.

**Theorem 14.** *For any $C$, if $K = \omega\left(\frac{N}{C}\right)$, that is the limit $\lim_{N\to\infty} \frac{N}{C \cdot K} = 0$, then*

$$\lim_{N\to\infty} J(X) = 1,$$

*so the allocation becomes fair as $N$ goes to $\infty$.*

*Proof.* The theorem follows from the fact that

$$\lim_{N\to\infty} J(X) = \lim_{N\to\infty} \frac{C \cdot K}{C \cdot K + (N - C)}$$

For this limit to be equal to 1, we need

$$\lim_{N\to\infty} \frac{N-C}{C \cdot K} = 0$$

that holds exactly when $K = \omega\left(\frac{N}{C}\right)$ (that is $K$ grows asymptotically faster than $\frac{N}{C}$; note that we assume that $C \leq N$). $\qquad\square$

For practical purposes, we may also need to know how big shall we choose $K$ given $C$ and $N$. The following theorem shows that:

**Theorem 15.** *Let $\varepsilon > 0$. If*

$$K > \frac{1 - \varepsilon}{\varepsilon}\left(\frac{N}{C} - 1\right),$$

*then $J(X) > 1 - \varepsilon$.*

*Proof.* The theorem follows straightforwardly from Equation 2. $\qquad\square$

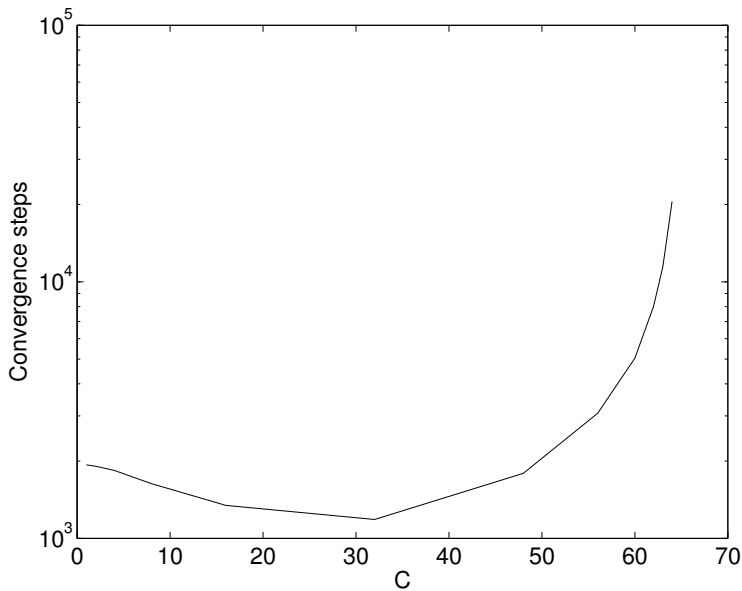

Figure 1: Average number of steps to convergence for $N = 64$, $K = N$ and $C \in \{1, 2, \ldots, N\}$.

## 6. Experimental Results

In all our experiments, we report average values over 128 runs of the same experiment. Error bars in the graphs denote the interval which contains the true expected value with probability 95%, provided that the samples follow normal distribution. The error bars are missing either when the graph reports values obtained theoretically (Jain index for the constant back-off scheme) or the confidence interval was too small for the scale of the graph.

### 6.1 Static Player Population

We will first analyze the case when the population of the players remains the same all the time.

#### 6.1.1 Convergence

First, we are interested in the convergence of our allocation algorithm. From Section 4 we know that it is polynomial. How many steps does the algorithm need to converge in practice?

Figure 1 presents the average number of convergence steps for $N = 64$, $K = N$ and increasing number of available channels $C \in \{1, 2, \ldots, N\}$. Interestingly, the convergence takes the longest time when $C = N$. The lowest convergence time is for $C = \frac{N}{2}$, and for $C = 1$ it increases again.

What happens when we change the size of the signal space $K$? Figure 2 shows the average number of steps to convergence for fixed $N$, $C$ and varying $K$. Theoretically, we have shown that the number convergence steps is $O(K \log K)$ in Theorem 12. However, in practice the convergence resembles linear dependency on $K$. This is because the algorithm needs to converge for all the coordination signals.

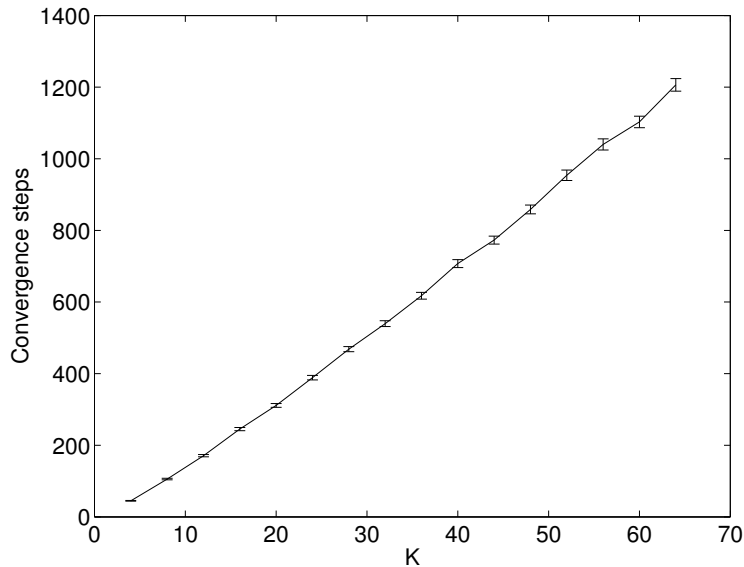

Figure 2: Average number of steps to convergence for $N = 64$, $C = \frac{N}{2}$ and $K \in \{2, \ldots, N\}$.

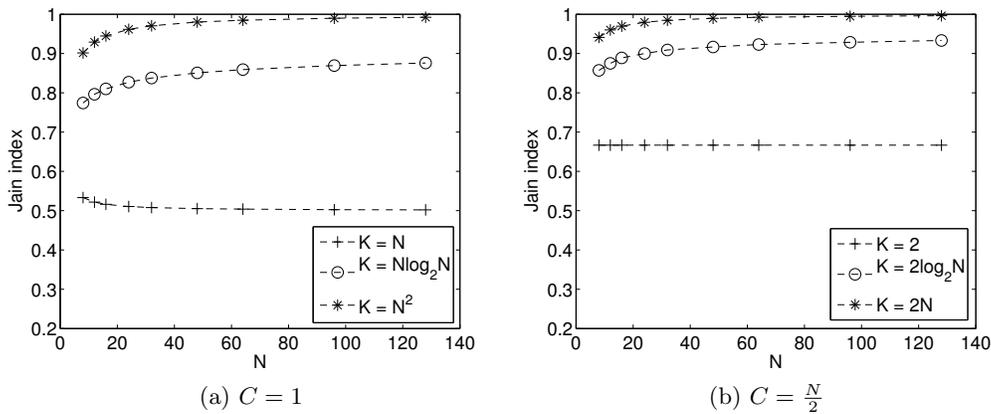

(a) $C = 1$                 (b) $C = \frac{N}{2}$

Figure 3: Jain fairness index for different settings of $C$ and $K$, for increasing $N$.

### 6.1.2 FAIRNESS

From Section 5, we know that when $K = \omega\left(\frac{N}{C}\right)$, the Jain fairness index converges to 1 as $N$ goes to infinity. But how fast is this convergence? How big do we need to choose $K$, depending on $N$ and $C$, to achieve a reasonable bound on fairness?

Figure 3 shows the Jain index as $N$ increases, for $C = 1$ and $C = \frac{N}{2}$ respectively, for various settings of $K$. Even though every time when $K = \omega\left(\frac{N}{C}\right)$ (that is, $K$ grows faster than $\frac{N}{C}$) the Jain index increases (as shown in Theorem 14), there is a marked difference between the various settings of $K$. When $K = \frac{N}{C}$, the Jain index is (from Equation 2):

$$J(X) = \frac{N}{2N - C}. \tag{3}$$

Therefore, for $C = 1$, the Jain index converges to 0.5, and for $C = \frac{N}{2}$, the Jain index is equal to $\frac{2}{3}$ for all $N > 0$, just as Figure 3 shows.

### 6.1.3 Optimizing Fairness

We saw how fair the outcome of the allocation algorithm is when agents consider the game for each signal value independently. However, is it the best we can do? Can we further improve the fairness, when each agent correlates her decisions for different signal values?

In a perfectly fair solution, every agent wins (and consequently can transmit) for the same number of signal values. However, we assume that agents do not know how many other agents there are in the system. Therefore, the agents do not know what is their fair share of signal values to transmit for. Nevertheless, they can still use the information in how many slots they already transmitted to decide whether they should back-off and stop transmitting when a collision occurs.

**Definition 8.** For a strategy $f_i^t$ of an agent $i$ in round $t$, we define its *cardinality* as the number of signals for which this strategy tells the agent to access:

$$|f_i^t| = \left|\left\{k \in \mathcal{K} | f_i^t(k) > 0\right\}\right|$$

Intuitively, agents whose strategies have higher cardinality should back-off more often than those with a strategy with low cardinality.

We compare the following variations of the channel allocation scheme, that differ from the original one only in the probability with which agents back off on collisions:

**Constant** The scheme described in Section 3; Every agent backs off with the same constant probability $p$.

**Linear** The back-off probability is $p = \frac{|f_i^t|}{K}$.

**Exponential** The back-off probability is $p = \mu^{\left(1 - \frac{|f_i^t|}{K}\right)}$ for some parameter $0 < \mu < 1$.

**Worst-agent-last** In case of a collision, the agent who has the *lowest* $|f_i^t|$ does not back off. The others who collided, do back off. This is a greedy algorithm that requires more information than what we assume that the agents have.

To compare the fairness of the allocations in experiments, we need to define the Jain index of an actual allocation. A resource allocation is a vector $\mathbb{X} = (X_1, X_2, \ldots, X_N)$, where $X_i$ is the cardinality of the strategy used by agent $i$. For an allocation $\mathbb{X}$, its Jain index is:

$$J(\mathbb{X}) = \frac{\left(\sum_{i=1}^{N} X_i\right)^2}{N \cdot \sum_{i=1}^{N} X_i^2}$$

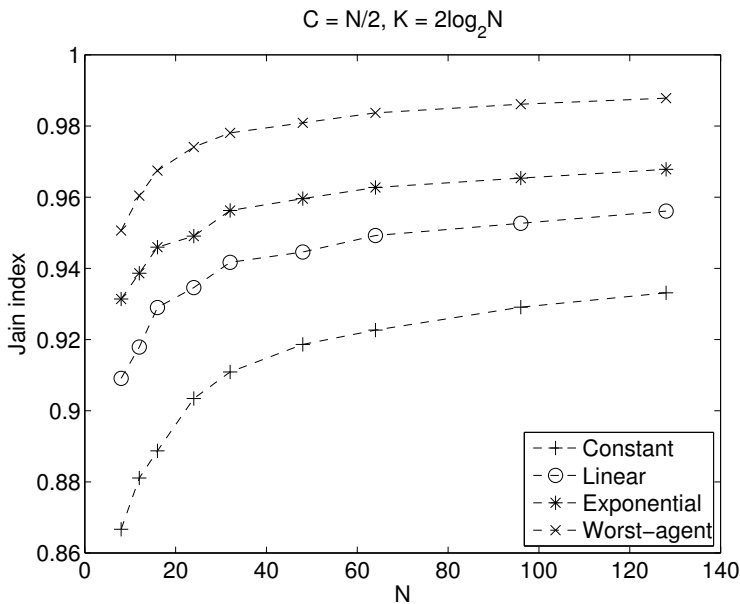

Figure 4: Jain fairness index of the channel allocation scheme for various back-off probabilities, $C = \frac{N}{2}$, $K = 2\log_2 N$

Figure 4 shows the average Jain fairness index of an allocation for the back-off probability variations. The fairness is approaching 1 for the *worst-agent-last* algorithm. It is the worst if everyone is using the same back-off probability. As the ratio between the back-off probability of the lowest-cardinality agent and the highest-cardinality agent decreases, the fairness increases.

This shows that we can improve fairness by using different back-off probabilities. Nevertheless, the shape of the fairness curve is the same for all of them. Furthermore, the exponential back off probabilities lead to much longer convergence, as shown on Figure 5. For $C = \frac{N}{2}$, the convergence time for the linear and constant back-off schemes is similar. The unrealistic *worst-agent-last* scheme is obviously the fastest, since it resolves collisions in 1 step, unlike the other back-off schemes.

## 6.2 Dynamic Player Population

Now we will take a look at the performance of our algorithm when the population of players is changing over time (either new players join or old players get replaced by new ones). We will also analyze the case when there are errors in what the players observe – either coordination signal or channel feedback is noisy.

### 6.2.1 JOINING PLAYERS

In this section, we will present the results of experiments where a group of players joins the system later. This corresponds to new nodes joining a wireless network. More precisely, 25% of the players join the network from the beginning. The remaining 75% of the players join the network later, one by one. A new player joins the network after the previous players have converged to a perfect channel allocation.

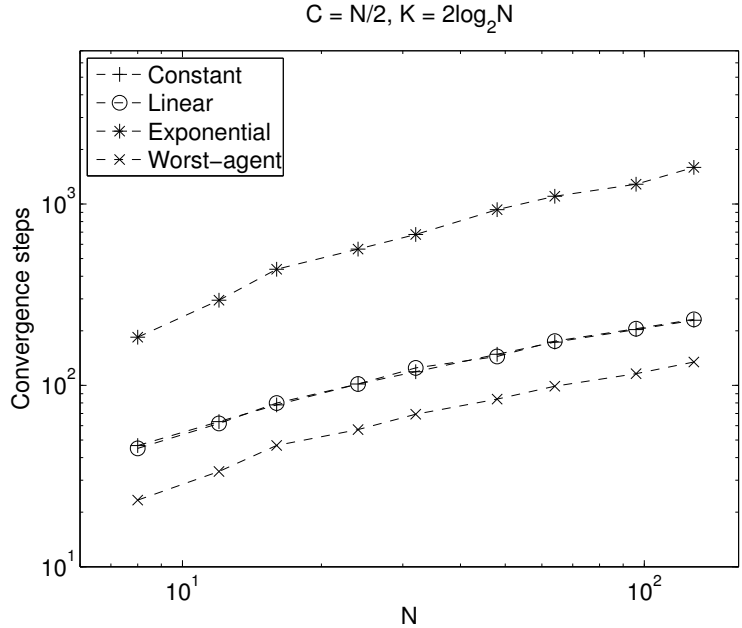

Figure 5: Convergence steps for various back-off probabilities.

We experiments with two ways of initializing a strategy of a new player.

**Greedy** Either, the joining players cannot observe how many other players there are already in the system. Therefore, their initial strategy tries to transmit in all possible slots.

**Polite** Or, players *do* observe $N(t)$, the number of other players who are already in the system at time $t$, when the new player joins the system. Therefore, their initial strategy tries to transmit in a slot only with probability $\frac{1}{N(t)}$.

Figure 6 shows the Jain index of the final allocation when 75% of the players join later, for $C = 1$. When the players who join are greedy, they are very aggressive. They start transmitting in all slots. On the other hand, if they are polite, they are not aggressive enough: A new player starts with a strategy that is as aggressive as the strategies of the players who are already in the system. The difference is that the new player will experience a collision in every slot she transmits in. The old players will only experience a collision in $\frac{1}{N(t)}$ of their slots. Therefore, they will back off in less slots.

Therefore, especially for the constant scheme, the resulting allocation is very unfair: either it is better for the new players (when they are greedy) or to the older players (when the players are polite).

This phenomenon is illustrated in Figure 7. It compares a measure called *group fairness*: the average throughput of the last 25% of players who joined the network at the end ("new" players) divided by the average throughput of the first 25% of players who join the network at the beginning ("old" players).

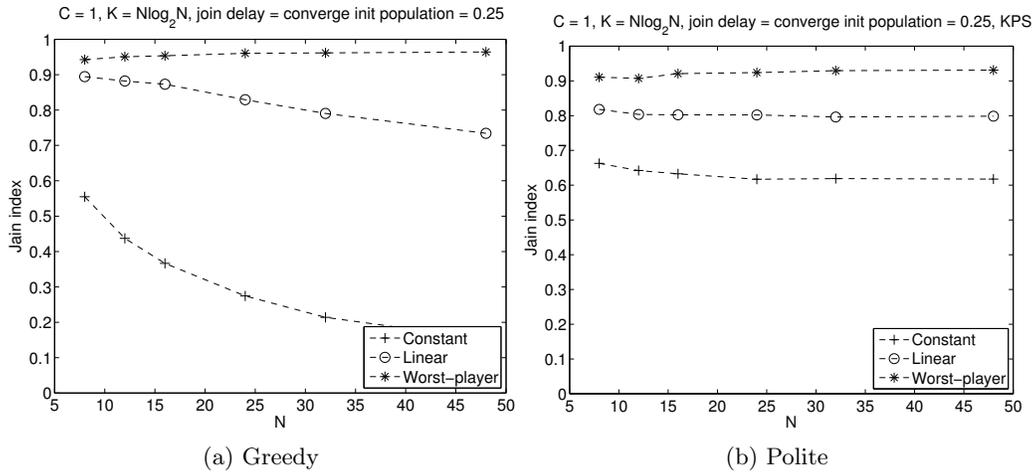

(a) Greedy  (b) Polite

Figure 6: Joining players, Jain index. $C = 1$ and $K = N \log_2 N$. The two graphs show the results for the two ways of initializing the strategy of a new player.

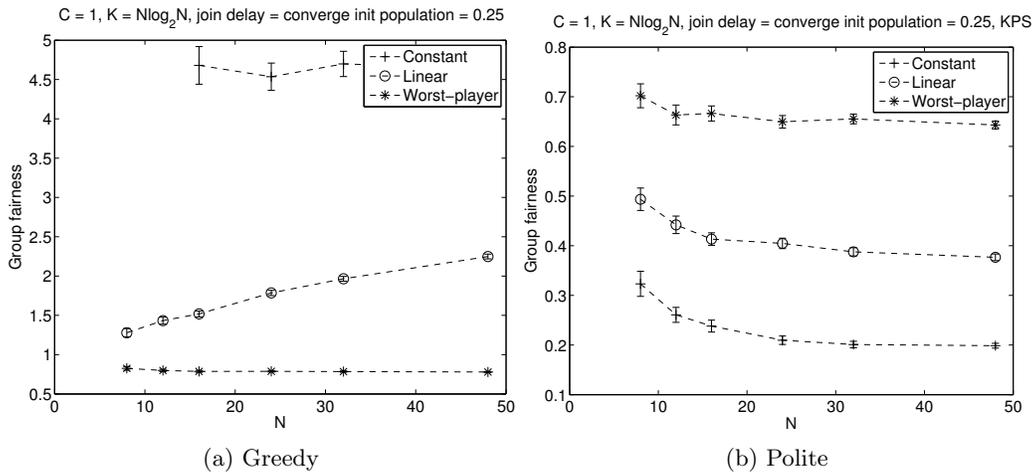

(a) Greedy  (b) Polite

Figure 7: Joining players, group fairness. $C = 1$ and $K = N \log_2 N$. The two graphs show the results for the two ways of initializing the strategy of a new player.

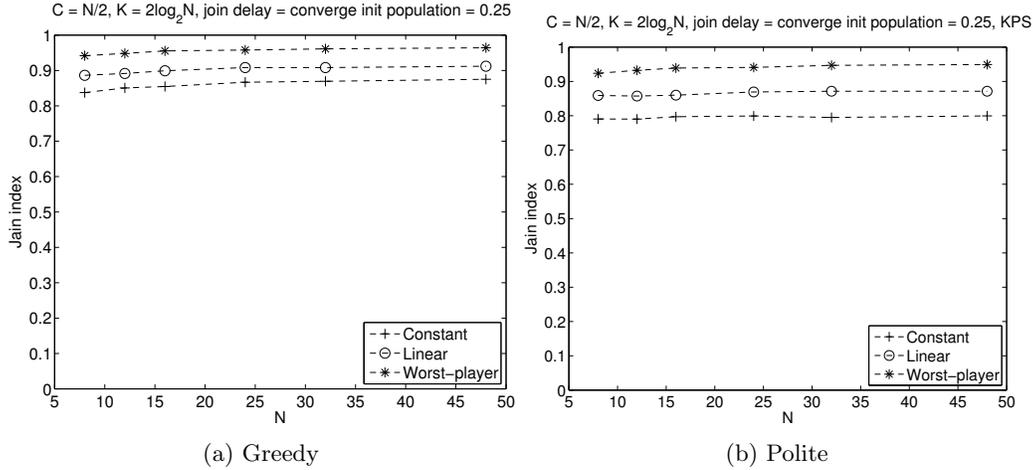

(a) Greedy

(b) Polite

Figure 8: Joining players, Jain index. $C = \frac{N}{2}$ and $K = 2\log_2 N$. The two graphs show the results for the two ways of initializing the strategy of a new player.

Let's look first at the case when the players are greedy. For the constant scheme, this ratio is around 4.5. For the linear scheme, this ratio is lower, although increasing as $N$ (the total number of players) grows. For the worst-player-last scheme, the ratio stays constant and interestingly, it is lower than 1, which means that "old" players are better off than "new" players.

When players are polite, this situation is opposite. Old players are way better off than new players. For the constant scheme, the throughput ratio is about 0.2.

Figures 8 and 9 show the same graphs for $C = \frac{N}{2}$. Here, the newly joining players are worse off even when they start transmitting in every slot. This is because while they experience a collision every time (because all channels in all slots are occupied), the old players only experience a collision with a probability $\frac{1}{\frac{N}{2}}$. On the other hand, the overall fairness of the whole population is better, because there are more channels to share and no agent can use more than one channel.

The difference between the old and new players is even more pronounced when the new players are polite.

### 6.2.2 Restarting Players

Another scenario we looked at was what happens when one of the old players "switches off" and is replaced with a new player with a randomly initialized strategy. We say that such a player got "restarted". In a wireless network, this corresponds to a situation when a user restarts their router. Note that the number of players in the network stays the same, it is just that some of the players forget what they have learned and start from scratch.

Specifically, in every round, for every player there is a probability $p_R$ that she will be restarted. After restart, she will start with a strategy that can be initialized in two ways:

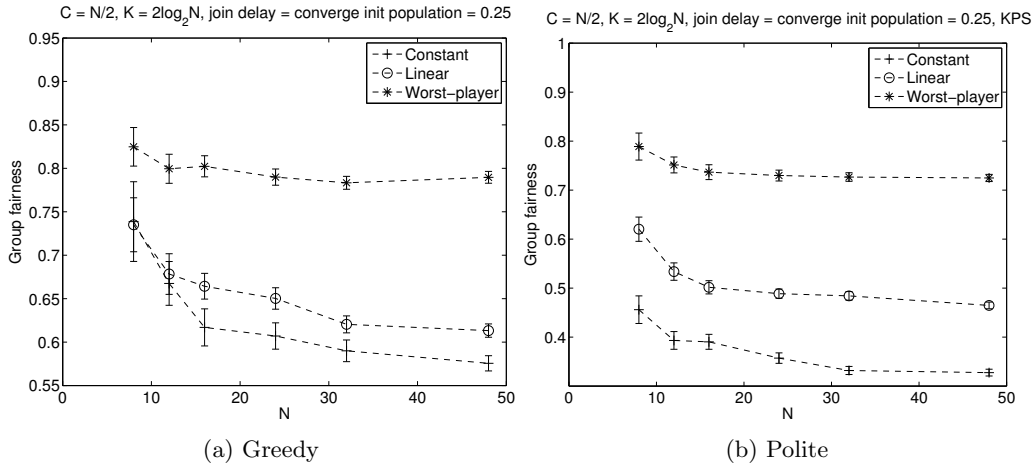

(a) Greedy

(b) Polite

Figure 9: Joining players, group fairness. $C = \frac{N}{2}$ and $K = 2\log_2 N$. The two graphs show the results for the two ways of initializing the strategy of a new player.

**Greedy** Assume that the player does not know $N$, the number of players in the system. Then for each signal value $k \in \mathcal{K}$ she chooses randomly $f_i(k) \in \mathcal{C}$. That means that she attempts to transmit in every slot on a randomly chosen channel.

**Polite** Assume the player *does* know $N$. For $k \in \mathcal{K}$, she chooses $f_i(k) \in \mathcal{C}$ with probability $\frac{C}{N}$, and $f_i(k) := 0$ otherwise.

Figure 10 shows the average overall throughput when $N = 32$, $C = 1$, and $K = N\log_2 N$ or $K = N$ for the two initialization schemes. A dotted line in all the four graphs shows the overall performance when players attempt to transmit in a randomly chosen channel with probability $\frac{C}{N}$. This baseline solution reaches $\frac{1}{e} \approx 37\%$ average throughput.

As the probability of restart increases, the average throughput decreases. When players get restarted and they are greedy, they attempt to transmit in every slot. If there is only one channel available, this means that such a restarted player causes a collision in every slot. Therefore, it is not surprising that when the restart probability $p_R = 10^{-1}$ and $N = 32$, the throughput is virtually 0: In every step, in expectation at least one player will get restarted, so there will be a collision almost always.

There is an interesting "phase transition" that occurs when $p_R \approx 10^{-4}$ for $K = N\log_2 N$, and when $p_R \approx 10^{-3}$ for $K = N$. There, the performance is about the same as in the baseline random access scenario (that requires the players to know $N$ though). Similar phase transition occurs when players are polite, even though the resulting throughput is higher, since the restarted players are less "aggressive".

Yet another interesting, but not at all surprising, phenomenon is that while the "worst-player-last" scheme still achieves the highest throughput, the "constant" back off scheme is better than the "linear" back-off scheme. This is because for the average overall throughput, it only matters how fast are the players able to reach a perfect allocation after a disruption. The worst-player-last scheme is the fastest, since it resolves a collision in 1 step. The con-stant scheme with back-off probability $p = \frac{1}{2}$ is worse (see Theorem 12). The linear scheme is the slowest.

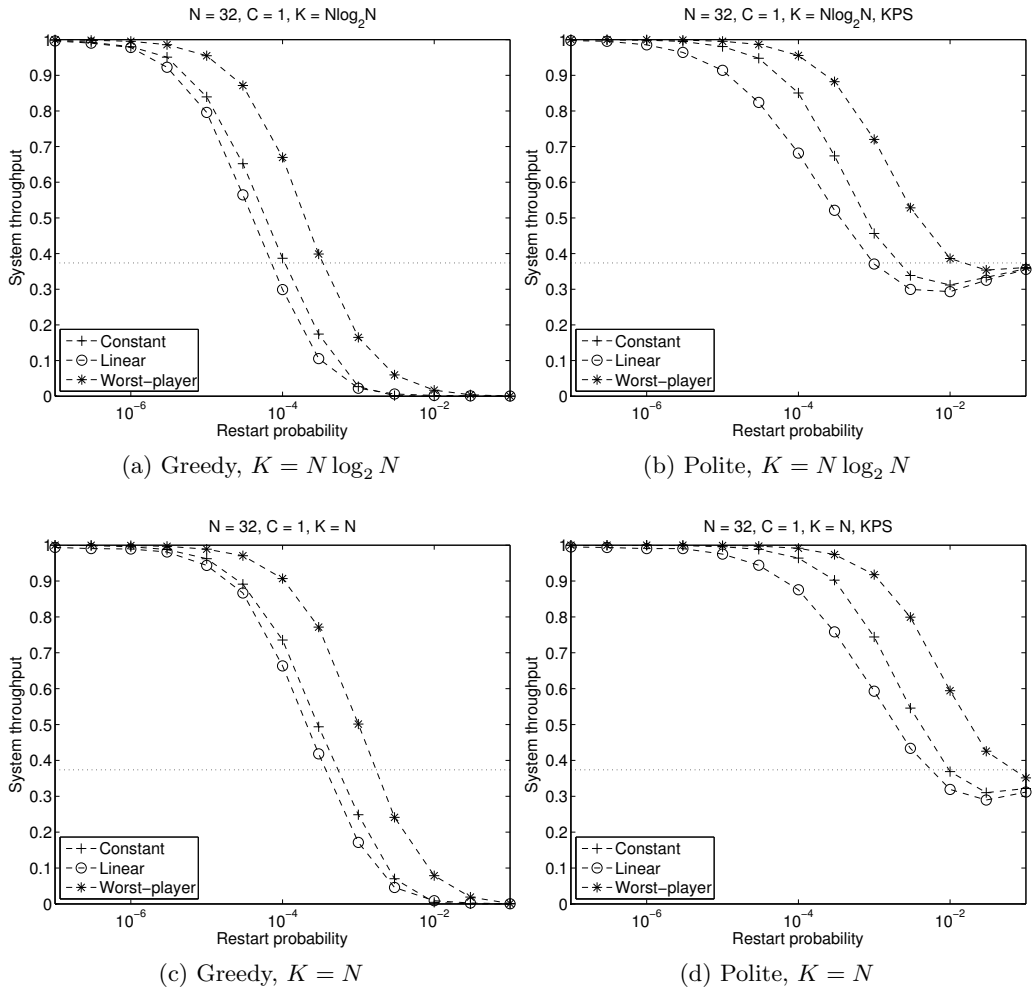

(a) Greedy, $K = N \log_2 N$

(b) Polite, $K = N \log_2 N$

(c) Greedy, $K = N$

(d) Polite, $K = N$

Figure 10: Restarting players, throughput, $N = 32$, $C = 1$

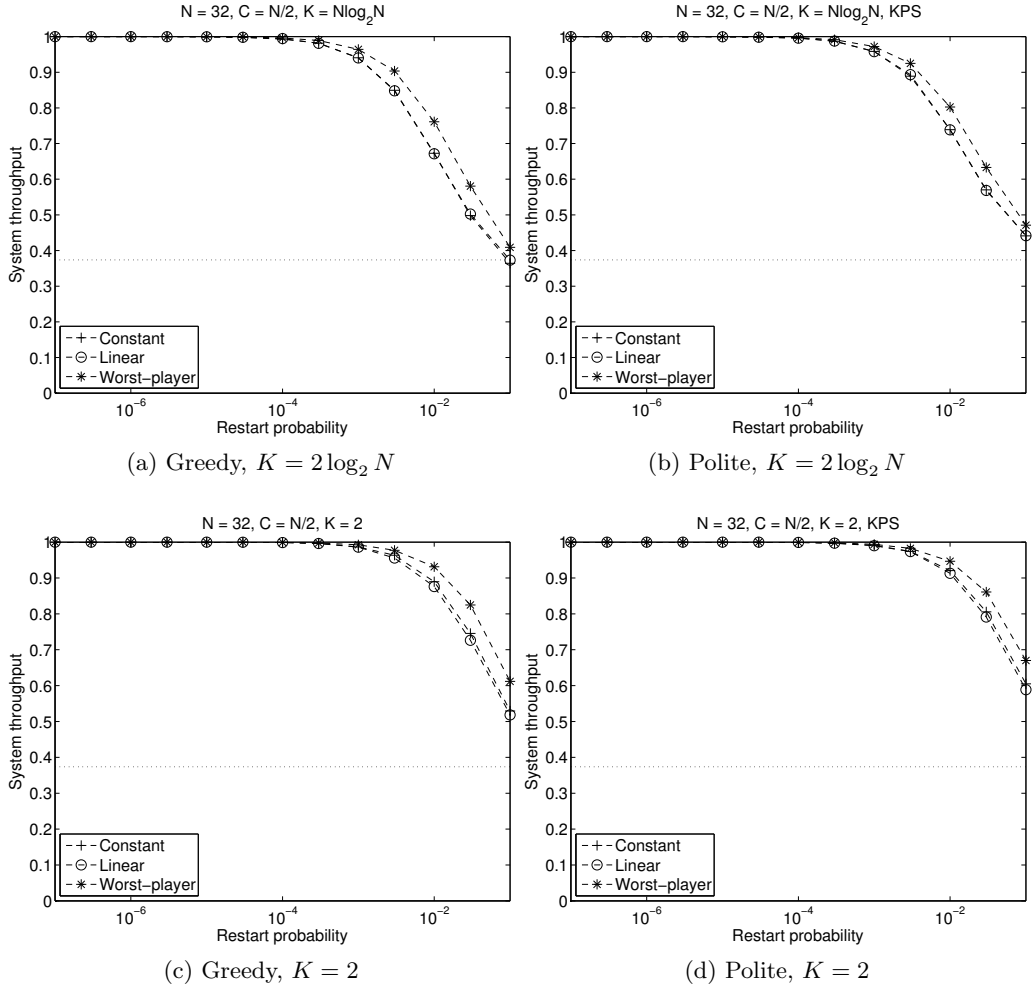

Figure 11: Restarting players, throughput, $N = 32$, $C = \frac{N}{2}$

Figure 11 shows the average overall throughput for $C = \frac{N}{2}$, and $K = \log_2 N$ or $K = 2$. There is no substantial difference between when players are greedy or polite. Since there are so many channels available, a restarted player will only cause a small number of collisions (in one channel out of $\frac{N}{2}$ in every slot), so the throughput will not decrease too much.

Also, the convergence time for linear and constant scheme is about the same when $C = \frac{N}{2}$, so they both adapt to the disruption equally well.

### 6.2.3 NOISY FEEDBACK

So far we assumed that players receive perfect feedback about whether their transmissions were successful or not. They could also observe the activity on a given channel perfectly. We are going to loosen this assumption now.

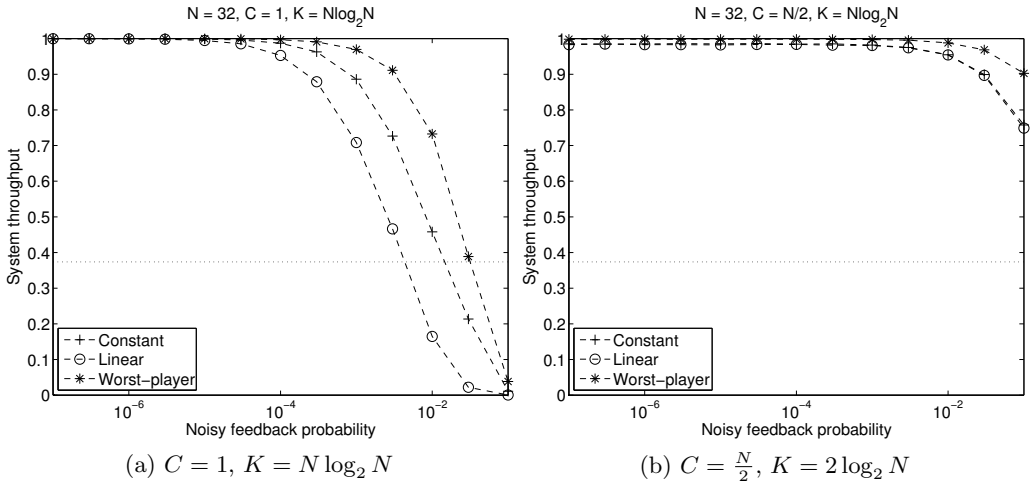

Figure 12: Noisy feedback, throughput, $N = 32$

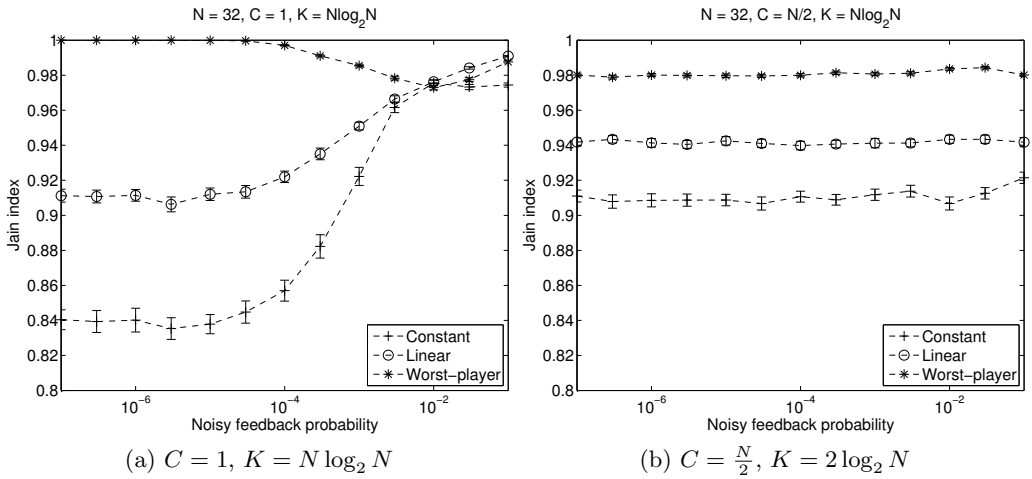

Figure 13: Noisy feedback, Jain index, $N = 32$

Suppose that in every step, every player has a probability $p_F$ that the feedback she receives was wrong. That is, if the player transmitted, she will learn that the transmission was successful when it was not, and vice versa. If the player observed some channel, she will learn that the channel was free when in fact it was not (and vice versa). In the context of wireless networks, this corresponds to an interference on the wireless channel.

How does this affect the learning?

In Figure 12 we show the average overall throughput when $C = 1$ and $C = \frac{N}{2}$ respectively. For one channel, the constant scheme is better than the linear scheme, because it adapts faster to disruptions. For $C = \frac{N}{2}$, both schemes are equivalent, because they are equally fast to adapt. A phase transition occurs when the noisy feedback probability is about $p_F = 10^{-2}$.

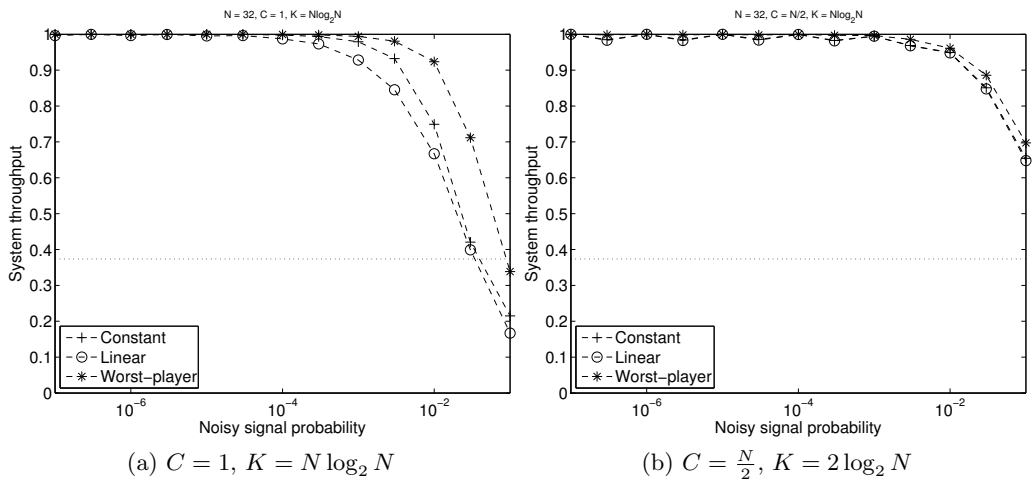

(a) $C = 1$, $K = N \log_2 N$           (b) $C = \frac{N}{2}$, $K = 2 \log_2 N$

Figure 14: Noisy coordination signal, throughput, $N = 32$

Figure 13 shows the Jain index of the allocation when players receive noisy feedback. As usual, the linear scheme is better than the constant, even though its throuput is lower (as we have shown above). Only when the overall throughput drops close to 0, all the schemes obviously have almost the same fairness.

### 6.2.4 NOISY COORDINATION SIGNAL

Our algorithm assumes that all players can observe the same coordination signal in every step. But where does this signal come from? It may be some random noise on a given frequency, an FM radio transmission etc. However, the coordination signal might be noisy, and different players can observe a different value. This means that their learning would be "out of sync". In the wireless networks, this corresponds to *clock drift*.

To see what happens in such a case, we use the following experiment. In every step, every player observes the correct signal (i.e. the one that is observed by everyone else) with probability $1 - p_S$. With probability $p_S$ she observes some other false signal (that is still taken uniformly at random from the set $\{0, ..., K - 1\}$).

The overall throughput is shown in Figure 14. We can see that the system is able to cope with a fairly high level of noise in the signal, and the drop in throughput only occurs as $p_S = 10^{-1}$. As was the case for experiments with noisy feedback, the constant back-off scheme is able to achieve a higher throughput thanks to its faster convergence.

The Jain index of the allocation (Figure 15) stays almost constant, only when the throughput drops the Jain index increases. When the allocation is more random, it is also more fair.

## 6.3 Generic Multi-agent Learning Algorithms

Several algorithms that are proved to converge to a correlated equilibrium have been proposed in the multi-agent learning literature. In the Introduction, we have mentioned three such learning algorithms (Foster & Vohra, 1997; Hart & Mas-Colell, 2000; Blum & Man-sour, 2007). However, the analysis of Foster and Vohra was only applicable to games of two players. In this section, we will briefly recall the other two multi-agent learning algorithms (Hart & Mas-Colell, 2000; Blum & Mansour, 2007), and compare their performance with our algorithm presented in Section 3.

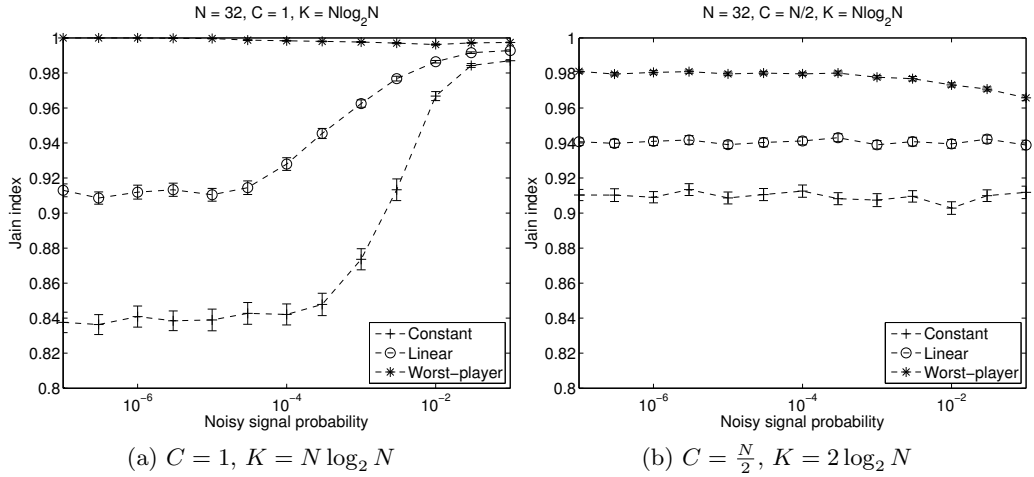

(a) $C = 1$, $K = N \log_2 N$          (b) $C = \frac{N}{2}$, $K = 2 \log_2 N$

Figure 15: Noisy coordination signal, Jain index, $N = 32$

The two algorithms we will compare our algorithm to are based on the notion of minimizing *regret* the agents experience from adopting a certain strategy. Intuitively, we can describe the concept of regret as follows: Imagine that an agent uses strategy $\sigma$ in a couple of rounds of the game, and accumulates a certain payoff. We would like to know how does this payoff compare to a payoff acquired by some simple alternative strategy $\tau$. The difference in the payoff between the strategy $\tau$ and $\sigma$ is the regret the agent perceives (ex-post) for choosing strategy $\sigma$ over strategy $\tau$.

What do we mean by "simple strategy"? One class of simple strategies are strategies that always select the same action. The *external regret* compares the performance of the strategy $\sigma$ to the performance of the best single action ex-post.

Another class of alternative strategies are strategies that modify strategy $\sigma$ slightly. Every time the strategy $\sigma$ proposes to play action $a$, the alternative strategy $\tau$ proposes action $a' \neq a$ instead. The *internal regret* is defined as the regret of strategy $\sigma$ compared to the best such alternative strategy. When all the agents adopt a strategy with low internal regret, they converge to a strategy profile that is close to a correlated equilibrium (also shown in Blum & Mansour, 2007).

Hart and Mas-Colell (2000) present a simple multi-agent learning algorithm that is guaranteed to converge to a correlated equilibrium. They assume that the players can observe the actions of all their opponents in every round of the game. Players start by choosing their actions randomly. Then they update their strategy as follows: Let $a_i$ be the action that player $i$ played in round $t-1$. For each action $a_j \in \mathcal{A}_i$, $a_j \neq a_i$, player $i$ calculates the difference between the average payoff she would have received had she played action $a_j$ instead of $a_i$ in the past, and the average payoff she received so far while playing action $a_i$. As we mentioned above, we can call this difference the *internal regret* of playing action $a_i$ instead of action $a_j$. The player then chooses the action to play in round $t$ with probability proportional to its internal regret compared to the previous action $a_i$. Actions with negative regret are never played. The previous action $a_i$ is played with positive probability – this way, the strategy has a certain inertia.

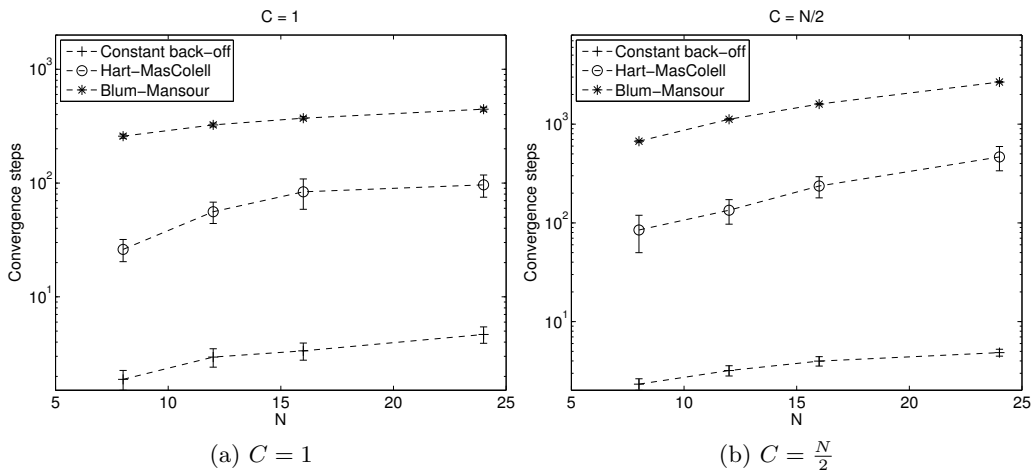

(a) $C = 1$

(b) $C = \frac{N}{2}$

Figure 16: General multi-agent learning algorithms, convergence rate.

Hart and Mas-Colell (2000) prove that if the agents adopt the adaptive procedure described above, the empirical distribution of the play (the relative frequency of playing a certain pure strategy profile) converges almost surely to the set of correlated equilibria.

Blum and Mansour (2007) present a general technique to convert any learning algorithm with low external regret to an algorithm with a low internal regret. The idea is to run multiple copies of the external regret algorithm. In each step, each copy returns a probability vector of playing each action. These probability vectors are then combined into one joint probability vector. When the player observes the payoff of playing each action, she updates the payoff beliefs of each external regret algorithms proportionally to the weight they had in the joint probability vector. The authors then show that when the players all use a learning algorithm with low internal regret, the empirical distribution of the game converges close to a correlated equilibrium.

One of the low-external-regret algorithms that Blum and Mansour (2007) present is the *Polynomial Weights (PW)* algorithm. There, a player keeps a weight for each of her actions. In every round of the game, she updates the weight proportionally to the loss (negative payoff) that action incurred in that round. Actions with higher weight get then chosen with a higher probability.

We have implemented the two generic multi-agent learning algorithms: The internal-regret-based algorithm of Hart and Mas-Colell (2000), and the PW algorithm of Blum and Mansour (2007). In all our experiments, both algorithms always converge to a pure-strategy Nash equilibrium of the channel allocation game, and therefore to an efficient allocation. However, the resulting allocation is not fair, as only a subset of agents of size $C$ can ever access the channels.

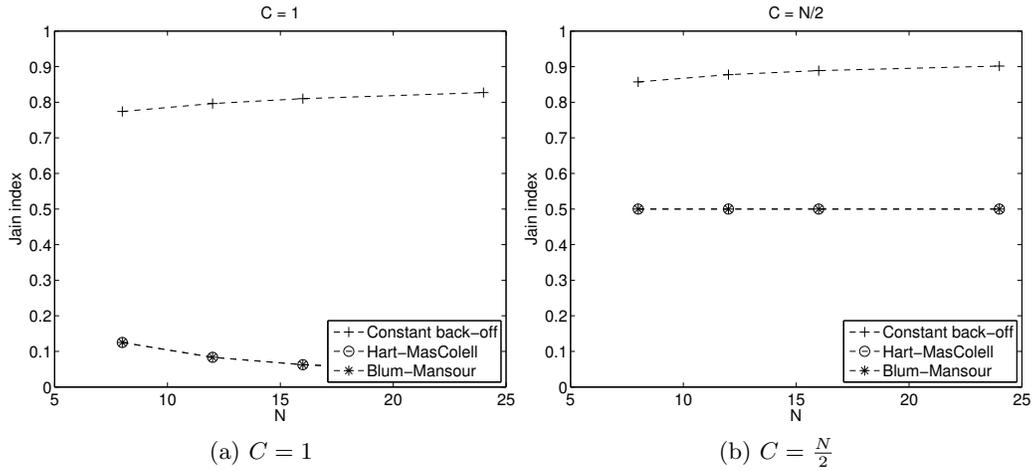

Figure 17: General multi-agent learning algorithms, Jain index.

Figure 16 shows the average number of rounds the algorithms take to converge to a stable outcome. We compare their performance with our learning algorithm from Section 3. For our learning algorithm, we set $K = 1$, so that it also only converges to a pure-strategy Nash equilibrium of the game. We performed 128 runs of each algorithm for each scenario. The error-bars in Figure 16 show the 95% confidence interval of the average, assuming that the convergence times are distributed according to a normal distribution.

Not surprisingly, the generic algorithms of Hart and Mas-Colell (2000) and Blum and Mansour (2007) cannot match the convergence speed of our algorithm, designed specifically for the problem of channel allocation. As the generic algorithms converge to a pure-strategy NE, the outcome is very unfair, and the Jain index is very low, as evidenced by Figure 17. We don't report the confidence bounds for the Jain index, as in all of the experiments the resulting Jain index was the same.

## 7. Related Work

Broadly speaking, in this paper we are interested in games where the payoff an agent receives from a certain action is inversely proportional to the number of other agents who chose the same action. How can we achieve efficient and fair outcome in such games? Variants of this problem have been studied in several previous works.

The simplest such variant is the *Minority game* (Challet, Marsili, & Zhang, 2005). In this game, $N$ agents have to simultaneously choose between two actions. Agents who chose an action that was chosen by a minority of agents receive a payoff of 1, whereas agents whose action choice was in majority receive a payoff of 0. This game has many pure-strategy Nash equilibria, where some group of $\lfloor \frac{N-1}{2} \rfloor$ agents chooses one action and the rest choose the other action. Such equilibria are efficient, since the largest possible number of agents achieve the maximum payoff. However, they are not fair: the payoff to the losing group of agents is always 0. This game has also one mixed-strategy NE that is fair: every agent chooses its action randomly. This equilibrium, on the other hand, is not efficient: the expected size of the minority group is lower than $\left\lfloor \frac{N-1}{2} \right\rfloor$ due to variance of the action selection.

Savit, Manuca, and Riolo (1999) show that if the agents receive feedback on which action was in the minority, they can learn to coordinate better to achieve a more efficient outcome in a repeated minority game. They do this by basing the agents' decisions on the history of past iterations. Cavagna (1999) shows that the same result can be achieved when agents base their decisions on the value of some random coordination signal instead of using the history. This is a direct inspiration for the idea of global coordination signal presented in this paper.

The ideas from the literature on Minority games have recently found their way into the cognitive radio literature. Mahonen and Petrova (2008) present a channel allocation problem much like ours. The agents learn which channel they should use using a strategy similar to the strategies for minority games. The difference is that instead of preferring the action chosen by the minority, in the channel allocation problem, an agent prefers channels which were not chosen by anyone else. Using this approach, Mahonen and Petrova are able to achieve a stable throughput of about 50% even when the number of agents who try to transmit over a channel increases. However, each agent is essentially choosing one out of a fixed set of strategies, that they cannot adapt. Therefore, it is very difficult to achieve a perfectly efficient channel allocation.

Wang et al. (2011) have implemented the algorithm from this work in an actual wireless network. In their setting, wireless devices are able to monitor the activity on all the channels. As a coordination signal, they have used the actual data packets that the agents send. The authors have shown that in practice, the learning algorithm (which they call *attachment learning*) improves the throughput 2-3× over the random access slotted ALOHA protocol.

Another, more general variant of our problem, called *dispersion game* was described by Grenager, Powers, and Shoham (2002). In a dispersion game, agents can choose from several actions, and they prefer the one that was chosen by the smallest number of agents. The authors define a *maximal dispersion outcome* as an outcome where no agent can move to an action with fewer agents. The set of maximal dispersion outcomes corresponds to the set of pure-strategy Nash equilibria of the game. They propose various strategies to converge to a maximal dispersion outcome, with different assumptions on the information available to the agents. On the contrary with our work, the individual agents in the dispersion games do not have any particular preference for the actions chosen or the equilibria which are achieved. Therefore, there are no issues with achieving a fair outcome.

Verbeeck, Nowé, Parent, and Tuyls (2007) use reinforcement learning, namely *linear reward-inaction automata*, to learn Nash equilibria in common and conflicting interest games. For the class of conflicting interest games (to which our channel allocation game belongs), they propose an algorithm that allows the agents to circulate between various pure-strategy Nash equilibria, so that the outcome of the game is fair. In contrast with our work, their solution requires more communication between agents, and it requires the agents to *know* when the strategies converged. In addition, linear reward-inaction automata are not guaranteed to converge to a pure-strategy NE in conflicting interest games; they may only converge to pure strategies.

All the games discussed above, including the channel allocation game, form part of the family of *potential games* introduced by Monderer and Shapley (1996). A game is called a potential game if it admits a *potential function*. A potential function is defined for every strategy profile, and quantifies the difference in payoffs when an agent unilaterally deviates from a given strategy profile. There are different kinds of potential functions: exact (where the difference in payoffs to the deviating agent corresponds directly to the difference in potential function), ordinal (where just the sign of the potential difference is the same as the sign of the payoff difference) etc.

Potential games have several nice properties. The most important is that any pure-strategy Nash equilibrium is just a local maximum of the potential function. For finite potential games, players can reach these equilibria by unilaterally playing the best-response, no matter what initial strategy profile they start from.

The existence of a natural learning algorithm to reach Nash equilibria makes potential games an interesting candidate for our future research. We would like to see to which kind of correlated equilibria can the agents converge there, and if they can use a simple correlation signal to coordinate.

## 8. Conclusions

In this paper, we proposed a new approach to reach efficient and fair solutions in multi-agent resource allocation problems. Instead of using a centralized, "smart" coordination device to compute the allocation, we use a "stupid" coordination signal, in general a random integer $k \in \{0, 1, \ldots, K-1\}$, that has no a priori relation to the problem. Agents then are "smart": they learn, for each value of the coordination signal, which action they should take.

From a game-theoretic perspective, the ideal outcome of the game is a correlated equilibrium. Our results show that using a global coordination signal, agents can learn to play a convex combination of pure-strategy Nash equilibria, that is a correlated equilibrium.

We showed a learning strategy that, for a variant of a channel allocation game, converges in expected polynomial number of steps to an efficient correlated equilibrium. We also proved that this equilibrium becomes increasingly fair as $K$, the number of available synchronization signals, increases.

We confirmed both the fast convergence as well as increasing fairness with increasing $K$ experimentally. We also investigated the performance of our learning strategy in case the agent population is dynamic. When new agents join the population, our learning strategy is still able to learn an efficient allocation. However, the fairness of this allocation will depend on how greedy the initial strategies of the new agents are. When agents restart at random intervals, it becomes more important how fast a strategy converges. A simple strategy where everyone backs off from transmitting with a constant probability is able to achieve higher throughput than a more sophisticated strategy where the back-off probability depends on in how many slots an agent is already transmitting. We also showed experimentally that the learning strategy is robust against noise in both the coordination signal, as well as in the feedback the agents receive about channel use. In both of these noisy scenarios, faster convergence of the constant back-off scheme helped achieve a higher throughput than the more fair linear back-off scheme. Finally, we compared the performance of out learning strategy with the generic multi-agent learning algorithms based on regret-minimization (Hart & Mas-Colell, 2000; Blum & Mansour, 2007). While these generic algorithms are theoretically proven to converge to a distribution of play which is close to a correlated equilibrium, they are not guaranteed to converge to a specific CE. Indeed, in our experiments, the algorithms of Hart and Mas-Colell and Blum and Mansour always converged to the efficient but unfair pure-strategy Nash equilibrium of the channel allocation game.

The learning algorithm presented in this paper has been implemented in a real wireless network by Wang et al. (2011), who have shown that it achieves 2-3× higher throughput than random access protocols such as ALOHA.

In this paper, we did not address the issue of whether non-cooperative but rational agents would follow the protocol we outlined. In our other work (Cigler & Faltings, 2012), we address this issue and show that under certain conditions, the protocol can be implemented in Nash equilibrium strategies of the infinitely repeated resource allocation game.